\documentclass[reprint,amsmath,amssymb,aps,twocolumn]{revtex4-1}
\usepackage{graphicx}
\usepackage{dcolumn}
\usepackage{rotating}
\usepackage{amsthm,amsmath,amssymb,amsfonts,bbm}
\usepackage{mathptmx}
\usepackage{mathrsfs}
\usepackage{xcolor}
\usepackage{siunitx}
\usepackage{bm}
\usepackage{textcomp}
\usepackage{url}
 
\begin{document}

\title{Electrical transport of surface-bound electrons on annealed solid neon}

\author{Ka Wing Yip} 
\email[E-mail:]{kawing.yip@oist.jp}
\affiliation{Quantum Dynamics Unit, Okinawa Institute of Science and Technology (OIST) Graduate University, Onna, 904-0495 Okinawa, Japan}
\author{Tomoyuki Tani} 
\altaffiliation[Current address: ]{Department of Applied Physics, Hokkaido University, Sapporo, 060-8628 Hokkaido, Japan}
\author{Jui-Yin Lin} 
\altaffiliation[Current address: ]{Fluid Mechanics Unit, Okinawa Institute of Science and Technology (OIST) Graduate University, Onna, 904-0495 Okinawa, Japan}
\affiliation{Quantum Dynamics Unit, Okinawa Institute of Science and Technology (OIST) Graduate University, Onna, 904-0495 Okinawa, Japan}
\author{Mikhail Belianchikov} 
\affiliation{Quantum Dynamics Unit, Okinawa Institute of Science and Technology (OIST) Graduate University, Onna, 904-0495 Okinawa, Japan}
\author{Denis Konstantinov}
\email[E-mail:]{denis@oist.jp}
\affiliation{Quantum Dynamics Unit, Okinawa Institute of Science and Technology (OIST) Graduate University, Onna, 904-0495 Okinawa, Japan}

\date{\today}


\begin{abstract}
We study the dc transport of electrons bound to the surface of solid neon using the conventional Sommer-Tanner method. First, we find that the capacitive cross-talk for the thickness of solid samples on the order of a millimeter varies substantially with the temperature and is strongly affected by the sample annealing. Second, we find that, even for the annealed samples, the observed response of electrons differs significantly from that of the surface-bound electrons on liquid helium and cannot be satisfactorily described by conventional models. Despite this, we managed to extract the resistance of the electron system from its in-phase response and investigate the temperature dependence in a wide range between 1.4  and 20 K. In a high-temperature range between 16 and 20 K, the resistance behavior is found to agree very well with the vapor-atom scattering mechanism, while at lower temperatures it is consistent with temperature-dependent scattering by surface roughness. Our analysis indicates a high electron mobility of electrons on solid neon but two orders of magnitude lower density than expected. These results suggest surface electrons as a suitable probe for studying the morphology of the neon surface but call for a better understanding of the experimental method applicable to such systems.   
\end{abstract}

\maketitle

\section{Introduction}
\label{intro}  

Electrons on solid neon are emerging as a promising platform for quantum information processing~\cite{JinNat2022,JinNatPhys2023,JinNatElec2026}. Bound to the surface of a non-polar dielectric by the image potential, with a binding energy of about 17~meV, such electrons reside in an exceptionally clean environment that ensures long coherence of their motional and spin states~\cite{Chen2022,Zou2026}. This system is reminiscent of the mobile electrons hovering over the surface of liquid helium, although the surface roughness of solid neon significantly reduces mobility of electrons and might even lead to their trapping~\cite{TrojJETP1979,TrojJETP1981,KanaiPRL2024}. The morphology of the neon surface is of significant interest in understanding the formation and control of the charge states of electrons used for quantum computing, and it has recently been investigated using superconducting resonators~\cite{MurchPRL2025,Matk2025,MurchPRAppl2026,Dizd2026}. Alternatively, mobile surface-bound electrons might be a useful probe to investigate the surface morphology by studying their electrical transport affected by the scattering of electrons due to the surface roughness~\cite{LeiderJLTP2025rew}. The dc transport of electrons on solid neon and hydrogen films has been studied several decades ago by the conventional capacitive-coupling (Sommer-Tanner) method~\cite{KajitaSS1984,KajitaSS1986}, which was originally developed for the electron-on-helium system~\cite{SommerPRL1971tanner}. The method can be described, with varying degree of sophistication, by a variety of models ranging from a simplified lumped-element circuit~\cite{IyeJLTP1980}, transmission-line model~\cite{DahmJLTP1987}, analytical solution of the Poisson equation~\cite{KovdJLTP1993}, and numerical simulations using the Green's function method~\cite{WilenJLTP1988}. Such models account well for the observed behavior of the mobile electrons on liquid helium and allow for the extraction of useful information about this system, such as the scattering mechanism. The density of surface electrons $n_\textrm{s}$, which are typically produced by thermionic emission from a tungsten filament placed above liquid helium, can be reliably determined by assuming complete screening of the applied electric field above the surface. Then, the electron mobility $\mu$ can be found from the measured resistance $R\propto (en_\textrm{s}\mu)^{-1}$ of the surface electrons ($e>0$ is the elementary charge). It is usually assumed that the same analysis holds for the observed behavior of electrons on solid neon and hydrogen. At low temperatures, the electron mobility below 10~m$^2$/Vs was found for surface electrons on solid samples prepared from the liquid phase by cooling it below the triple point temperature $T_\textrm{t}$ (24.56~K and 13.8~K for neon and hydrogen, respectively)~\cite{TrojJETP1981,EdelJLTP1983,KajitaSS1984,IyeSS1986}. This mobility can be attributed to the scattering by surface roughness~\cite{SokolovPRB1995}. Significantly lower mobility was found for solid samples prepared by quench condensation of hydrogen vapor below the triple point, but it could be significantly enhanced by subsequent annealing of solid samples~\cite{Kono1JLTP1991}.

In the described studies, the measurement results are typically presented in terms of the values of the resistance $R$ and mobility $\mu$ extracted from the data based on the assumption that the conventional models used for description of electrons on helium are also applied for the case of the solid substrates. Motivated by a recent interest in surface-bound electrons on solid neon as a promising platform for quantum technological applications, we investigated the dc transport of such electrons using a capacitive Corbino setup in a wide temperature range between 1.4 and 20~K. We found that, contrary to the expected similarity of this system to a system of mobile electrons on liquid helium, an electron response differs significantly from that of electrons on helium. The conventional models fail to account for both in-phase and quadrature components of the response of the system, as they usually do for electrons on helium. Nevertheless, we managed to extract the resistance $R$ of the electron system from the in-phase response and investigate its temperature dependence. In a range of high temperatures between 16 and 20~K, the variation of $R$ is found to agree well with the scattering of electrons by neon vapor atoms. At lower temperatures, the behavior is consistent with a temperature-dependent scattering mechanism due to surface roughness. We find that the density of electrons over solid neon is significantly lower than that expected from the condition of complete screening of the electric field above the surface. This implies the formation of a localized charge inside the bulk solid neon due to incident electrons penetrating the surface barrier. Our analysis, which relies on the calculated mobility of electrons due to the scattering by neon vapor atoms, indicates the electron mobility due to the surface-roughness scattering that significantly exceeds $10$~m$^2$/Vs, which was reported earlier. This substantial difference suggests that further studies are warranted.  

\section{Setup and method}
\label{sec:setup}

\begin{figure*}[htp]
\includegraphics[width=\textwidth,keepaspectratio]{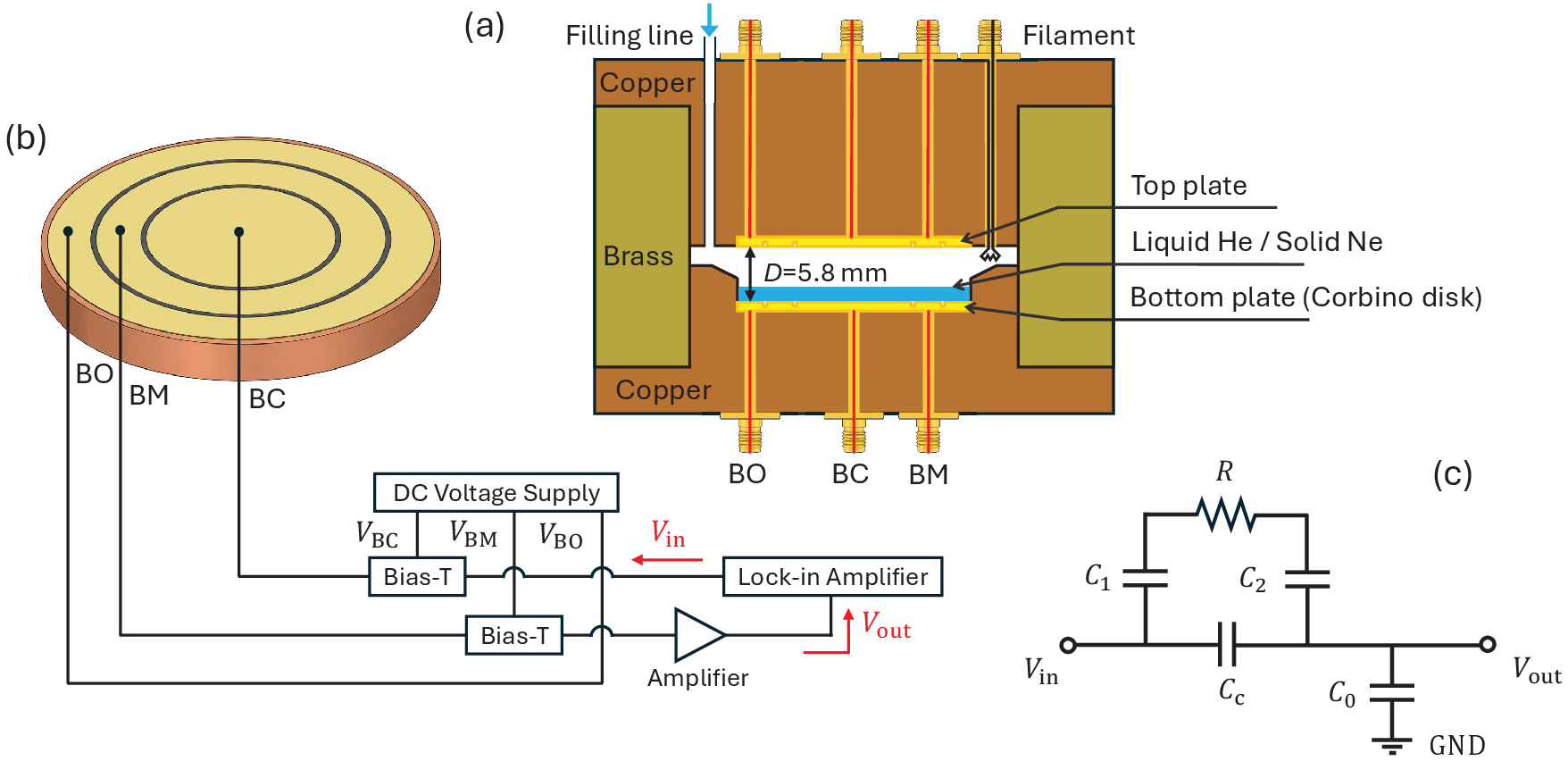} 
\caption{\label{fig:1} (color online) (a) Schematic drawing of the experimental cell. (b) 3D rendering of the Corbino disk and block diagram of the measurement setup. (c) Equivalent electrical lumped-element circuit of the Sommer-Tanner measurement setup.}
\end{figure*}

The schematic drawing of the experimental setup is shown in Fig.~1. Our leak-tight experimental cell is attached to the cold plate of a continuous-flow 1~K refrigerator (ICEoxford 1.0K Dry Systems) and consists of two parts made of copper separated by a middle part made of brass (see Fig.~\ref{fig:1}(a)). The upper and lower parts contain circular conducting plates separated by a distance $D=5.8~\mathrm{mm}$, which form a parallel-plate capacitor. Each plate, having a diameter of 26~mm, forms a Corbino disk for conductivity measurements consisting of three concentric electrodes separated by two circular gaps (width 0.2~mm) with diameters 14 and 20~mm (see Fig.~\ref{fig:1}(b)). Independent dc bias voltages $V_\textrm{BC}$, $V_\textrm{BM}$ and $V_\textrm{BO}$ can be applied to the three electrodes of the bottom Corbino disk, which we refer to as the center (BC), middle (BM) and outer (BO) electrodes, respectively. In the work described here, the three electrodes of the top Corbino disk are dc grounded. Helium or neon gas is supplied to the cell through a capillary filling line with 0.6~mm inner diameter and condensed on top of the bottom Corbino disk by adjusting the temperature of the cell. For neon condensation, the temperature is maintained at about 25~K, that is above the triple-point temperature, and subsequently reduced to the lowest temperature of about 1.2~K for neon solidification. More details of the preparation of solid neon are given in Sec.~\ref{sec:anneal}. For helium condensation, the temperature is maintained at the lowest temperature of 1.2~K. The thickness $d$ of the liquid-helium or solid-neon sample was estimated from the measured capacitance $C_\textrm{BT}$ between the central electrodes of the top and bottom Corbino disks according to

\begin{equation}
d=\frac{\epsilon D}{\epsilon-1}\left(1-\frac{C_\mathrm{BT}^{(0)}}{C_\mathrm{BT}}\right),
\label{eq:d}
\end{equation}

\noindent where $\epsilon$ is the dielectric constant of a sample and $C_\mathrm{BT}^{(0)}$ is the capacitance measured for an empty cell.

The surface of a sample is charged with electrons at 1.2~K using the thermionic emission from a tungsten filament mounted near the top Corbino disk (see Fig.~\ref{fig:1}(a)), while a positive voltage $V$ is applied to all three electrodes of the bottom Corbino disk. Conventionally, the density of surface electrons can be estimated from the complete screening of the electric field above the surface according to    

\begin{equation}
n_{\mathrm s}=\frac{\epsilon_0\epsilon V}{ed},
\end{equation}

\noindent where $\epsilon_0$ is the electrical permittivity of vacuum. After charging the surface, the voltage $V_\textrm{BO}$ applied to the outer electrode was usually swept to a negative value of -32~V to confine electrons above the central and middle electrodes of the Corbino disk. 

The transport of surface electrons is measured by the conventional Sommer-Tanner method using the Corbino disk of the bottom plate. A small ac voltage signal $V_\textrm{in}$ of angular frequency $\omega$ is applied to the central electrode of the disk, in addition to the dc bias voltage, thus exciting the radial motion of surface electrons, while an electrical current $I$ induced by this motion at the middle electrode is detected as a voltage drop at the capacitive load $C_0$ of the connecting cable according to $V_\textrm{out}=I/(i\omega C_0)$. This voltage signal is amplified by a room-temperature amplifier and detected using a lock-in amplifier referenced at the driving frequency $\omega$. Conventionally, the ac response of the system is analyzed by an equivalent lumped-element circuit shown in Fig.~\ref{fig:1}(c)~\cite{IyeJLTP1980}. Here, $C_1$ ($C_2$) represents the capacitance between the surface charge and the central (middle) electrode of the Corbino disk, while $C_\textrm{c}$ accounts for the direct coupling between the central and middle electrodes. The resistance $R$ represents the resistance of the surface-charge sheet. It relates to the electric conductivity of electrons $\sigma$ by $R=\gamma/\sigma$, where $\gamma$ is the geometric factor that represents the ratio of the length of a current sheet to its width. From an elementary circuit analysis, we can write the normalized voltage signal detected by the lock-in amplifier as

\begin{equation}
\frac{V_\textrm{out}}{V_\textrm{in}} = \frac{G}{C_0}\left( C_\textrm{c} + \frac{C}{1+(\omega C R)^2} - \frac{i\omega C^2R}{1+(\omega C R)^2} \right),
\label{eq:vout}
\end{equation}

\noindent where $G$ is the gain of the amplifier and $C^{-1}=C_1^{-1}+C_2^{-1}$. The first term on the right-hand side of Eq.~\eqref{eq:vout} represents the cross-talk voltage signal due to the direct capacitive coupling between the electrodes of the Corbino disk. It is important to note that, in general, the coupling capacitance $C_\textrm{c}$ depends on the dielectric sample that covers the Corbino disk. The value of $C_\textrm{c}=1.932$~pF was measured directly at 1.2~K for an empty cell using a capacitance bridge. From this value, the capacitive load presented by the cable was determined to be $C_0=164.674$~pF from the measured cross-talk signal. The last two terms on the right-hand side of Eq.~\eqref{eq:vout} correspond to the in-phase and quadrature components of the surface-electron response. In the limit $R\rightarrow 0$, the quadrature component vanishes, while the in-phase component represents a purely capacitive response of the electron system, corresponding to the maximum voltage signal due to the electrons given by $V_0=V_\textrm{in}G(C/C_0)$. In an opposite limit $R\rightarrow \infty$, the electron response vanishes.   

\section{Response of electrons on liquid helium}
\label{sec:eHe}

\begin{figure}[htp]
\includegraphics[width=\columnwidth,keepaspectratio]{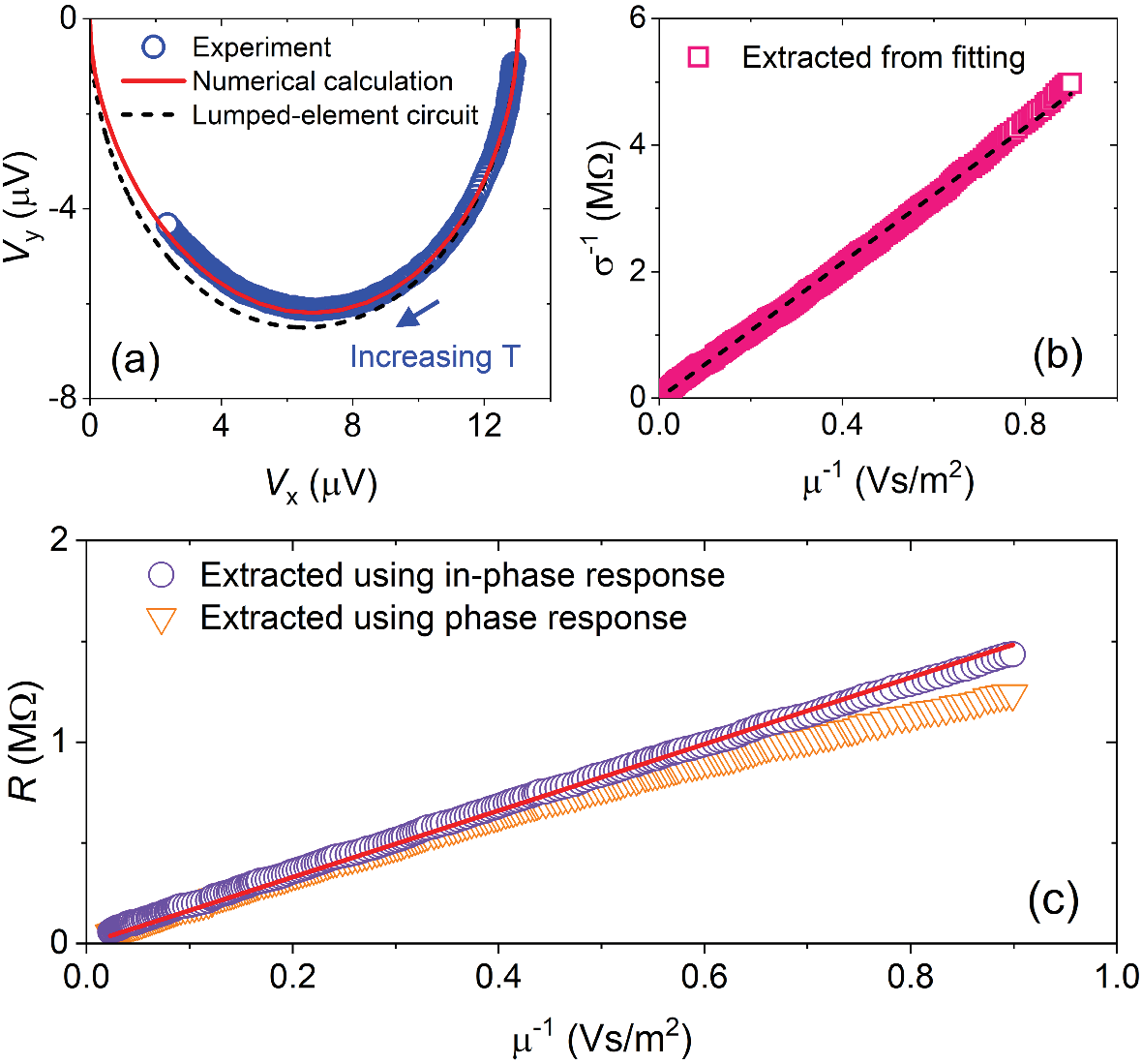} 
\caption{\label{fig:2} (color online) (a) The Argand plot of the measured Sommer-Tanner response $V_\textrm{out}$ of electrons on liquid helium-4 obtained by varying the temperature of the experimental cell between $T=1.17$ and 2.3~K (circles). Solid line shows a fitted response calculated by solving the electrodynamic problem of the electric field distribution inside the cell, while the dashed line shows the calculated response using the lumped element $RC$ circuit, as described in Sec.~\ref{sec:setup}. (b) The inverse conductivity $\sigma^{-1}$ (squares) extracted from the measured response by the numerical fitting (solid line in panel (a)) versus the inverse mobility calculated for the scattering of electrons by the helium-vapor atoms. Dashed line is a linear fit to $\sigma^{-1}$ versus $\mu^{-1}$. (c) The resistance $R$ extracted from the in-phase component (circles) and the phase (triangles) of the measured response $V_\textrm{out}$ versus the inverse mobility calculated for the scattering of electrons by the helium-vapor atoms. The solid line is a linear fit to the experimental data extracted from the in-phase response. }
\end{figure}

To make a direct comparison between electron response on liquid helium, as described by the conventional methods, and that on solid neon using the same experimental setup, the transport measurements were first conducted on superfluid helium. Helium-4 is condensed into the experimental cell and the level of liquid helium is set at a distance $d=1.3$~mm determined from the capacitance measurement, as described earlier. The liquid surface is charged by heating the filament, with a positive bias voltage $V=32$~V applied to all the bottom electrodes. The density of surface electrons estimated from the electric-field screening is $n_\textrm{s}=\epsilon_0\epsilon V/(ed)= 1.4\times 10^{12}$~m$^{-2}$, where $\epsilon=1.056$ is the dielectric constant of liquid helium. After deposition, the bias voltage at the outer electrode of the Corbino disk is set at -32~V. The electron response is measured by applying an ac voltage signal $V_\textrm{in}=10$~mV$_\mathrm{rms}$ at a frequency $\omega/(2\pi)=1.1$~MHz to the central electrode of the Corbino disk, with the output amplifier gain set to unity. Fig.~\ref{fig:2}(a) shows the Argand plot of the measured in-phase ($V_x$) and quadrature ($V_y$) components of the output voltage $V_\textrm{out}$ as the temperature varies between 1.17 and 2.17~K, the superfluid transition temperature of helium-4. The constant cross-talk voltage signal arising from the direct coupling between the BC and BM electrodes was subtracted from the data. In the considered temperature range, the mobility of electrons is expected to be dominated by the scattering of helium vapor atoms and decreases exponentially with $T$~\cite{SaitohJPSJ1977}. At the lowest $T$, the mobility of electrons $\mu$ is high and the response is predominantly capacitive. As mobility drops with increasing temperature, the Argant plot passes through the locus and the amplitude of the output signal decreases towards zero. The relation between the measured signal and the electron conductivity $\sigma = e n_\mathrm{s}\mu$ of electrons can be found by solving the electrodynamic problem of the electric field distribution inside the cell in the presence of surface charge. Assuming an azimuthal symmetry of the Corbino geometry employed in our experiment, this problem can be analytically solved by expanding in a series of the Bessel functions~\cite{KovdJLTP1993,SyvokLTP1996}. The calculated output signal for varying values of $\sigma$ is plotted by a solid line in Fig.~\ref{fig:2}(a). In our calculations, the diameter of the circular pool of electrons is used as an adjustable parameter to match the maximum capacitive response $V_0$ at the low-temperature limit. The agreement of the calculated response with the measured one is very good. A similar result is obtained by solving the above problem using the Green's function method~\cite{WilenJLTP1988}. By comparing the calculated response with the measured one, the conductivity of the electron system $\sigma$ is established as a function of temperature $T$.   

To compare with the expected temperature behavior of $\sigma (T)$, we calculated the mobility of electrons due to the vapor-atom scattering according to $\mu=8eb/(3\pi\hbar A n_\textrm{G})$, where $b=7.6$~nm is the extension of the electronic wave function for the ground surface-bound state, $A=4.98\times 10^{-20}$~m$^2$ is the cross-section of electron scattering by the helium atom, and $n_\textrm{G}$ is the density of helium-gas atoms~\cite{SaitohJPSJ1977}. The latter is given by

\begin{equation}
n_\textrm{G} = \left(\frac{Mk_\textrm{B}T}{2\pi \hbar^2}\right)^{3/2} \exp \left( -\frac{Q}{k_\textrm{B}T} \right),
\end{equation}

\noindent where $M=6.64\times 10^{-27}$~kg is the mass of helium atom and $Q/k_\textrm{B}=7.17$~K is the vaporization energy. The inverse conductivity $\sigma^{-1}$, which is extracted from the experimental data, versus the calculated mobility is plotted by squares in Fig.~\ref{fig:2}(b). The temperature dependence of $\sigma^{-1}(T)$ is expected to follow that of the inverse mobility $\mu(T)^{-1}$, with a temperature-independent proportionality coefficient given by $(en_\textrm{s})^{-1}$. The expected linear relationship between $\sigma^{-1}$ and $\mu^{-1}$ is in good agreement with our result shown in Fig.~\ref{fig:2}. From a linear-dependence fit to the data (dashed line in Fig.~\ref{fig:2}(c)), we obtain the electron density $n_\textrm{s}=1.2\times 10^{12}$~m$^{-2}$, which is in good agreement with the earlier estimate from the electric-field screening.   

Alternatively, we can try to analyze the data using the simplified lumped-element circuit model discussed in Sec.~\ref{sec:setup}. The dashed line in Fig.~\ref{fig:2}(a) shows the electron response calculated using Eq.~\eqref{eq:vout} by varying the resistance $R$, with the parameter $V_0$ set at 13~$\mu$V to match the maximum capacitive response in the limit $R\rightarrow 0$. From this value, we can estimate the capacitance $C$ between the surface charge and Corbino disk according to $C=C_0(V_0/V_\textrm{in})=0.214$~pF. As can be seen from Fig.~\ref{fig:2}(a), the agreement of the lumped-element circuit model with the measured response is still satisfactory, although not as good as the numerical calculation given by the solid line. Conventionally, the resistance $R$ of the electron system is extracted from the phase $\phi$ of the complex normalized response given by Eq.~\eqref{eq:vout} according to $R=-\tan(\phi)/(\omega C)$~\cite{SommerPRL1971tanner}. However, anticipating difficulties with interpreting the transport data obtained for electrons on solid neon, as will be discussed in Sec~\ref{sec:eNe}, here we attempt to extract the resistance $R$ of the electron system using only the in-phase component $V_x$ of the measured voltage signal $V_\textrm{out}$. According to Eq.~\eqref{eq:vout}, the resistance $R$ can be expressed using the normalized in-phase response as
 
\begin{equation}
R=(\omega C)^{-1} \sqrt{\frac{V_0}{V_x}-1},
\label{eq:vx}
\end{equation}

\noindent The values of $R$ extracted from the experimental data in Fig.~\ref{fig:2}(a) using Eq.~\eqref{eq:vx} are plotted by circles in Fig.~\ref{fig:2}(c) versus the calculated inverse mobility. Our experimental result for $R$ is in good agreement with the expected linear relationship between $R=\gamma/(en_\textrm{s}\mu)$ and $\mu^{-1}$. The geometric factor $\gamma$ can be found from the proportionality coefficient $\gamma/(en_\textrm{s})$ given by the slope of the linear fit to $R$ (solid line in Fig.~\ref{fig:2}(c)). Using $n_\textrm{s}=1.2\times 10^{12}$~m$^{-2}$ found earlier, we obtain $\gamma\approx 0.29$. We note that if the electron density $n_\textrm{s}$ can be estimated independently, for example from the condition of the electric field screening, which is usually assumed, the mobility can be calculated from the measured values of $R$ (up to an unknown factor $\gamma$) and compared with the theory. Anticipating difficulties encountered with an accurate estimation of $n_\textrm{s}$ for electrons on solid neon, as described in Sec.~\ref{sec:eNe}, here we take an alternative approach of comparing the experimental values of $R$ with the theoretical values of mobility, from which $\gamma/n_\textrm{s}$ can be determined. 

Finally, for the sake of comparison, the resistance values extracted from the phase $\phi$ of the electron response according to the relation $R=-\tan(\phi)/(\omega C)$ are shown by the triangles in Fig.~\ref{fig:2}(c). This gives a similar result as for the in-phase response. To summarize the main results of this section, the analysis based on the in-phase response of the electron system in our Sommer-Tanner setup can provide an adequate description of the dc transport of the surface electrons on liquid helium. This method is used to analyze the transport of electrons on solid neon, as described in Sec.~\ref{sec:eNe}.      

\section{Annealing of solid neon}
\label{sec:anneal}

The analysis of temperature-dependent transport presented in Sec.~\ref{sec:eHe} assumes a temperature-independent cross-talk signal $V_\textrm{in}G(C_\textrm{c}/C_0)$ due to the direct capacitive coupling between electrodes of the Corbino disk. This cross-talk signal is subtracted from the temperature-dependent data to obtain the Argant plot in Fig.~\ref{fig:2}(a). However, the coupling capacitance $C_\textrm{c}$ can vary with temperature $T$ due to the corresponding variation in the density of the dielectric sample that covers the Corbino disk, both through the variation in the thickness of the sample and its dielectric constant. In case of liquid helium-4 considered in Sec.~\ref{sec:eHe}, the relative variation of the density $\rho$ in the range between 1.17 and 2.17~K is $\Delta \rho/\rho\approx 6\times 10^{-3}$, resulting in the capacitance variation not exceeding $\Delta C_\textrm{c}/C_\textrm{c}\approx 10^{-3}$~\cite{HarrisPRA1970}. Since $\Delta C_\textrm{c}<<C$, such a variation can be safely neglected in the analysis of the data for electrons on liquid helium. For solid neon samples, the relative variation of the density in the temperature range between 1.4 and 20 K is expected to be larger, $\Delta \rho/\rho\approx 3\times 10^{-2}$~\cite{AcoPRB2000}. Moreover, we find that the actual variation of the coupling capacitance $C_\textrm{c}$ with temperature could be significantly larger than expected from the above variation of density $\rho$, which can obstruct the extraction of the resistance $R(T)$ from the experimental data. Fortunately, it was found that the variation of $C_\textrm{c}$ with temperature could be greatly reduced by the annealing of solid neon. We describe these results in this Section.

\begin{figure}[htp]
\includegraphics[width=\columnwidth,keepaspectratio]{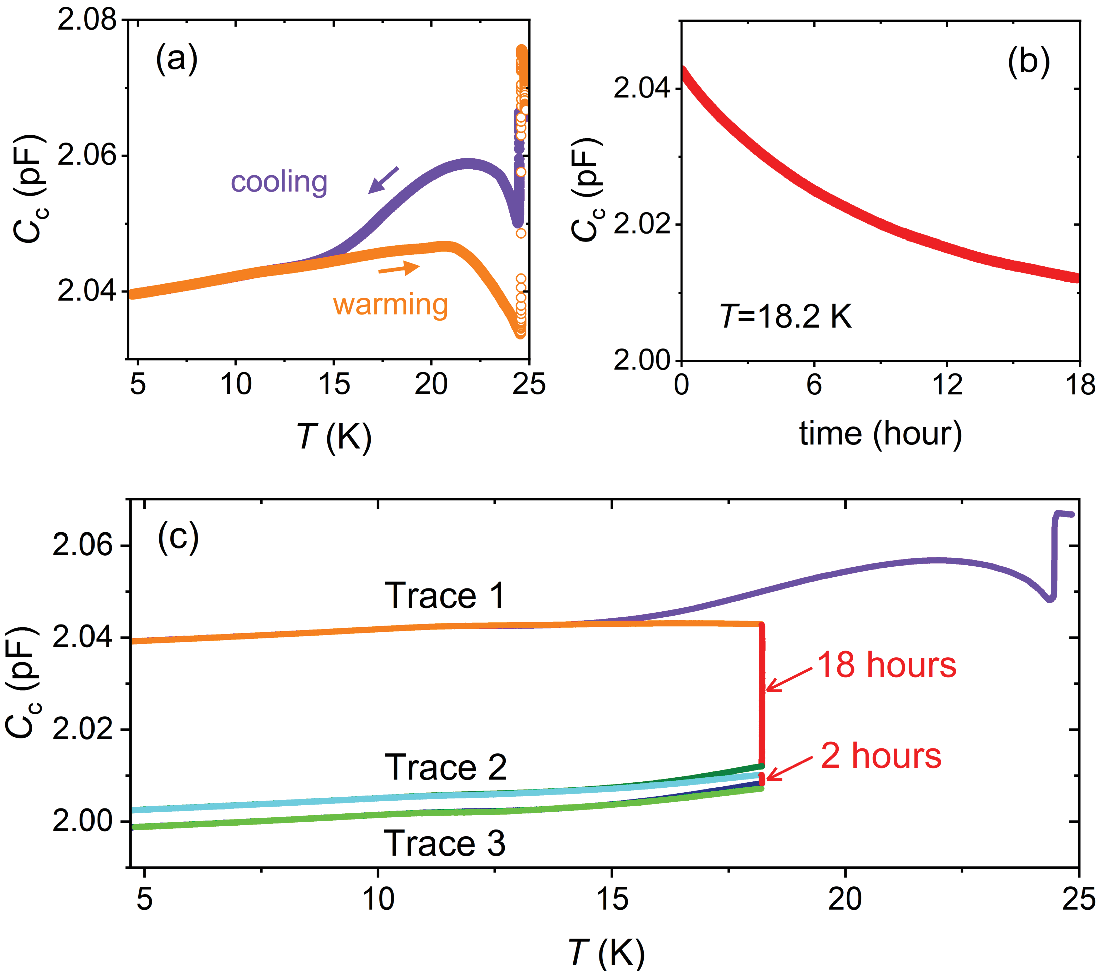} 
\caption{\label{fig:4} (color online) (a) Capacitance $C_\textrm{c}$ measured between the central and middle electrodes of the bottom Corbino disk covered by neon of thickness 1.23~mm during cooling the experimental cell from 25 to 4.75~K, followed by warming it back to 25 K. (b) Capacitance $C_\textrm{c}$ recorded during annealing of solid neon at a fixed temperature of the cell $T=18.2$~K. (c) Capacitance $C_\textrm{c}$ recorded during an initial cooling of neon from 25~K to 4.75~K and warming to 18.2~K (trace 1), followed by annealing of neon for 18 hours at a fixed temperature $T=18.2$~K, then followed by two more cooling and warming cycles (traces 2 and 3) separated by another annealing for 2 hours at $T=18.2$~K.}  
\end{figure}

Figure~\ref{fig:4}(a) shows an exemplary plot of $C_{\mathrm c}$ measured by a capacitance bridge as a function of cell temperature $T$ during initial cooling of liquid neon of thickness $d=1.23$~mm, determined from capacitance measurements (see Sec~\ref{sec:setup}), from 25 to 4.75~K, followed by its warming to 25~K, at a fixed rate of $3~\mathrm{K/hour}$. An abrupt decrease in $C_{\mathrm c}$ is observed near the triple-point temperature $T_\textrm{t}=24.56$~K, which corresponds to the neon crystallization. Below $13~\mathrm{K}$, the values of $C_{\mathrm c}$ measured during the cooling and warming of the sample are almost identical. However, at higher temperatures, the measured capacitance differs for the cooling and warming of the sample, with lower values of $C_\textrm{c}$ during the warming. Once the sample converts back to the liquid phase above the triple-point temperature, the capacitance is restored to its original value before cooling. From the observed behavior we conclude that, at sufficiently high temperatures, some structural changes occur in the bulk solid neon over time. This results in a thermal-history dependence of the cross-talk signal. As explained earlier, this dependence is detrimental to our measurements of $R(T)$ at high $T$. However, we find that annealing solid neon at high $T$ greatly reduces this unwanted dependence. First, we find that during annealing of solid neon at a fixed temperature, the capacitance decreases with time. This effect is illustrated in Fig.~\ref{fig:4}(b) where we shower a typical dependence of $C_\textrm{c}$ on time taken at $T=18.2$~K. After 18 hours of annealing, the capacitance decreases by about 1.5\% and continues to decrease at a progressively slower rate. It is found that such an annealing of the sample at high $T$ significantly reduces the thermal-history effect shown in Fig.~\ref{fig:4}(a). For example, Fig.~\ref{fig:4}(c) shows the thermal-history effect before and after annealing. Here, the neon sample was first cooled from 25~K to 4.75~K and then warmed to 18.2~K, after which it was annealed at this temperature for 18~hours, as shown in Fig.~\ref{fig:4}(b). After the capacitance dropped due to annealing, the cooling and warming cycle was repeated with the same cooling/warming rate (trace 2). Remarkably, the thermal-history effect is significantly reduced as a result of annealing. After the second cooling-warming cycle, the solid sample was annealed again for 2 hours, followed by the third cooling-warming cycle (trace 3). During the second annealing, the capacitance dropped by approximately 0.1\%, while the thermal-history effect was reduced even further. From this observation we conclude that sufficiently long annealing at high temperature removes the thermal-history effect almost completely. For such annealed samples, the measured capacitance $C_\textrm{c}$ is found to vary by about  $\Delta C_\textrm{c}/C_\textrm{c}\approx 7\times 10^{-3}$ in the temperature range from 1.4 to 20~K, which is in reasonable agreement with the corresponding variation of the density $\rho$ of solid neon, as discussed earlier.

An observed slow decrease in capacitance $C_\textrm{c}$ suggests that neon relaxes towards a stable solid structure while maintained at a sufficiently high temperature. Several effects could contribute to the change in $C_{\mathrm c}$, including changes in the density, strain, crystalline defects, and grain structure. Unfortunately, it is impossible to make any definite conclusion about such possibilities from our measurements. In particular, such measurements do not provide any evidence that annealing improves morphology of the solid neon surface. Nevertheless, such measurements are important to establish a time-independent capacitive cross-talk in a typical Sommer-Tanner setup to correctly interpret the conductivity measurements .

\section{Response of electrons on solid neon}
\label{sec:eNe} 

\begin{figure}[htp]
\includegraphics[width=\columnwidth,keepaspectratio]{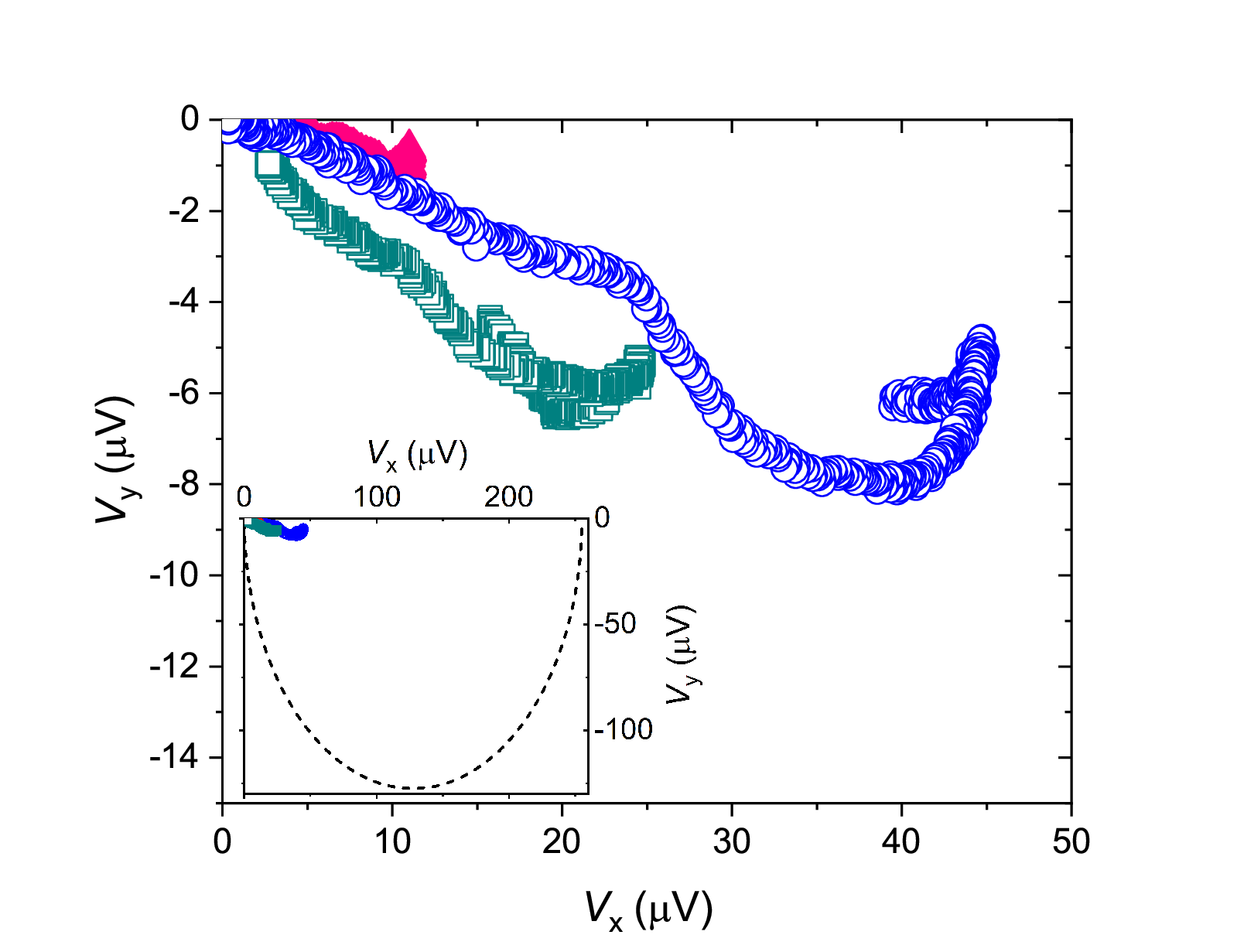} 
\caption{\label{fig:6} (color online) The Argand plot of the measured Sommer-Tanner response of the electron system on three different solid-neon samples measured by varying the temperature of the experimental cell between $T=1.4$ and 20~K. In the inset: The electron response predicted by the lumped-element model (dashed line). For comparison, the symbols show the same plots as on the main panel.}  
\end{figure}

After annealing solid neon to ensure that the Corbino disk showed a time-independent capacitive cross-talk signal at a fixed temperature, the temperature-dependent cross-talk signal was measured by warming the cell from 1.4 to 20 K. After that, it was cooled back to 1.4~K and charged with electrons as described in Sec.~\ref{sec:setup}, with the positive voltage $V=32$~V applied to all the bottom electrodes. As previously, the expected electron density from the electric field screening can be calculated as $n_\textrm{s}=\epsilon_0\epsilon V/(ed)$, where $\epsilon =1.244$ is the dielectric constant of solid neon and $d$ is the thickness of the solid sample. After the deposition, the electron response was measured by increasing the cell temperature to 20~K, and the measured cross-talk signal at a given $T$ was subtracted from the data at the corresponding temperature. Three different annealed samples of comparable thickness have been investigated and showed similar results. The open circles in Fig.~\ref{fig:6} show the electron response using a sample of thickness $d=0.78$~mm, which was determined from the capacitance measurements at low $T$. This particular sample was annealed at $T=20$~K for 36 hours until the measured capacitance $C_\textrm{c}$ reached a steady value to ensure a time-independent cross-talk signal at a fixed temperature in the range from 1.4 to 20 K. The expected electron density from the electric field screening is estimated to be $n_\textrm{s} =2.7\times 10^{12}$~m$^{-2}$. Electron transport for this sample was measured by applying an ac voltage signal $V_\textrm{in}=10$~mV$_\mathrm{rms}$ at a frequency of 750~kHz to the Corbino disk input, with the output amplifier gain set at $G=10$. As can be seen, the Argand plot of the measured response of electrons on solid neon significantly deviates from that of electrons on superfluid helium. Similar behavior was found for two other investigated samples, as shown by the open squares and solid triangles in Fig.~\ref{fig:6}.

\begin{figure}[htp]
\includegraphics[width=\columnwidth,keepaspectratio]{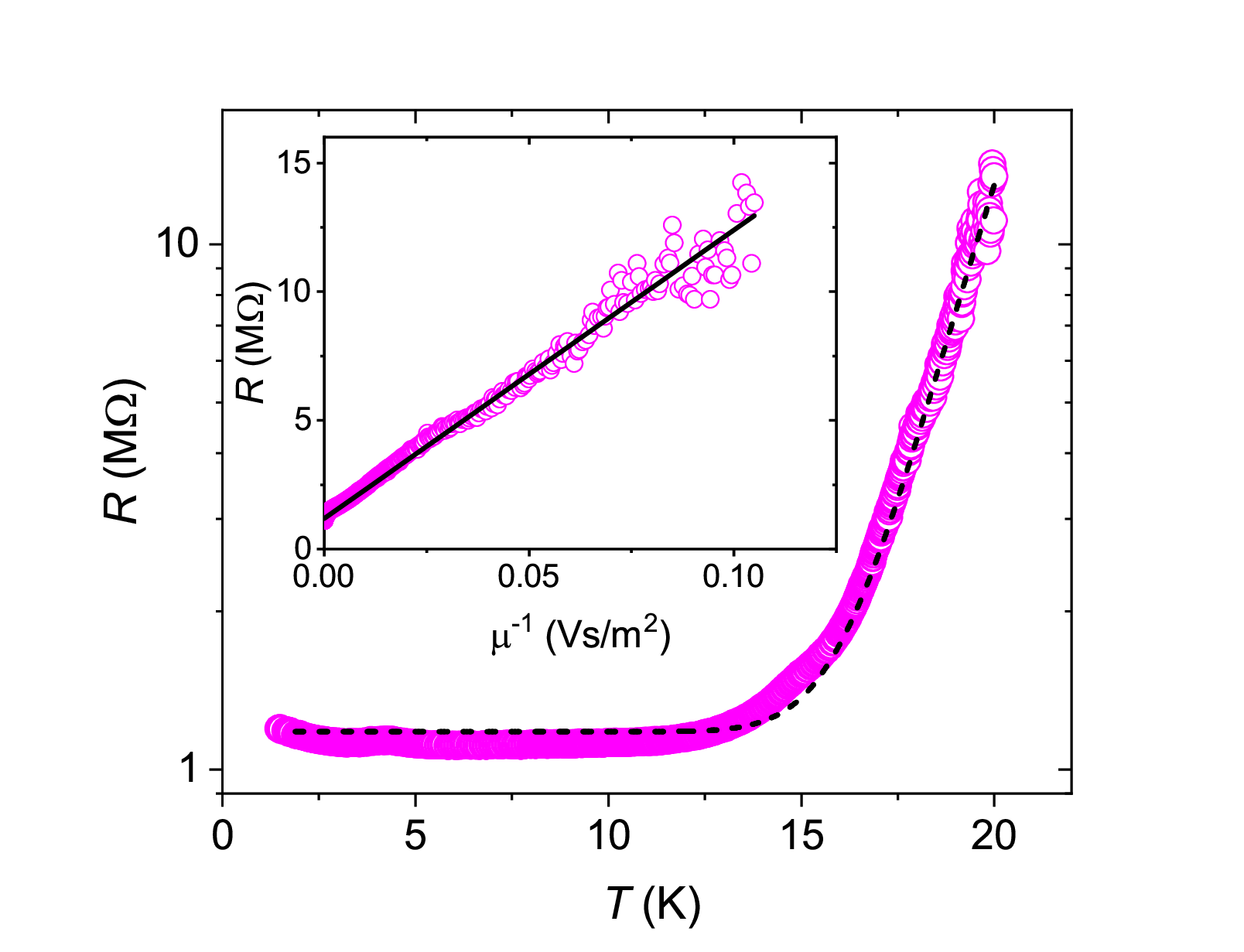} 
\caption{\label{fig:7} (color online) The resistance of electrons on solid neon extracted from the experimental data shown in Fig.~\ref{fig:6} versus temperature. The dashed line is a fit to Eq.\eqref{eq:R} using the calculated values of mobility due to the neon-vapor scattering. In the inset: The experimental values of $R$ versus the calculated inverse mobility $\mu^{-1}$. The solid line is a linear fit to Eq.\eqref{eq:R}.}  
\end{figure}

In what follows, we try to analyze the electron response using the lumped-element circuit, as described in Sec.~\ref{sec:setup}. First, we establish the maximum value of the capacitive response $V_0$ in the limit $R \rightarrow 0$ given by $V_0=V_\textrm{in}(C/C_0)$. Because the capacitance between the Corbino disk and the charge layer over the solid neon is not known, we estimate it using the earlier result for $C=0.214$~pF for electrons on superfluid helium by taking into account the ratio between the dielectric constants of solid neon and liquid helium of about 1.18, as well as the ratio between the thicknesses of the solid neon and liquid helium samples of about 1.67. This gives us an estimate $C=0.42$~pF, which corresponds to $V_0=V_\textrm{in}G(C/C_0)=255$~$\mu$V. The Argand plot of the expected electron response predicted by the lumped-element circuit is shown by the dashed line in the inset of Fig.~\ref{fig:6}. Clearly, unlike for the electron-on-helium system, an observed complex response of the electron system on solid neon significantly deviates from the model. Nevertheless, we can try to extract the resistance $R$ of the electron system from the in-phase component of the response using Eq.~\eqref{eq:vx}. Fig.~\ref{fig:7} shows a semi-log plot of $R$ versus temperature $T$ extracted from the measured response shown by the open circles in Fig.~\ref{fig:6}. In the high-temperature range between 16 and 20~K, the resistance shows an exponential dependence on $T$ that indicates the scattering of electrons by neon-vapor atoms. The corresponding mobility of electrons can be extracted from $R$ providing that the electron density $n_\textrm{s}$ and geometric factor $\gamma$ are known. Alternatively, we can compare the measured resistivity with the calculated values of $\mu$ to find the ratio $n_\textrm{s}/\gamma$, as was done in Sec.~\ref{sec:eHe}. We calculate the electron mobility using the same relation as that used for electrons on liquid helium, that is $\mu=8eb/(3\pi\hbar A n_\textrm{G})$, where $b=1.9$~nm is the extension of the electronic wave function for the ground surface-bound state, $A=4\pi a_\textrm{s}^2 = 1.815\times 10^{-20}$~m$^2$ is the cross-section of electron scattering by the neon atom ($a_\textrm{s}=0.38~\AA$ is the scattering length~\cite{EbnPRA1979}), and $n_\textrm{G}$ is the density of neon-gas atoms. In the considered temperature range from 16 to 20~K, which corresponds to the vapor pressure varying from 1.5 to 39~mbar, the vapor-atom density can be calculated using the ideal-gas approximation $n_\textrm{G}=P/(k_\textrm{B}T)$, where the sublimation vapor pressure of neon $P$ (in bar) can be calculated from the Antoine equation

\begin{equation}
P=10^{A-\frac{B}{T+C}},
\label{eq:Antoine}
\end{equation}

\noindent with the empirical coefficients given by $A=3.7564$, $B=95.599$ K, and $C=-1.503$ K~\cite{NIST}. Following Ref.~\cite{KajitaSS1984}, we can represent the total resistance $R$ of the electron system as a sum of the contributions from the scattering of electrons by the surface roughness and by neon-vapor atoms, as follows from the Matthiessen's rule for the corresponding scattering rates. Thus, we write

\begin{equation}
R = R_\textrm{S} + \frac{\gamma}{en_\textrm{s}\mu},
\label{eq:R}
\end{equation}

\noindent where $R_\textrm{S}$ is the resistance due to scattering of electrons by surface roughness. The measured resistance $R$ versus the calculated inverse mobility $\mu^{-1}$ due to neon-vapor scattering is shown in the inset of Fig.~\ref{fig:7}. From the linear fit to Eq.~\eqref{eq:R} shown by the solid line we determine $R_\textrm{s}=1.18$~M$\Omega$ and $n_\textrm{s}/\gamma=5.6\times 10^{10}$~m$^{-2}$. Despite the simplifying assumption that the contribution of the surface-roughness scattering is independent of $T$, the observed relationship between $R$ and $\mu^{-1}$ is almost linear, allowing its slope to be reliably determined. Using an earlier estimate for $\gamma=0.29$, we estimate the electron density $n_\textrm{s}=1.6\times 10^{10}$~m$^{-2}$. Surprisingly, this is significantly lower (more than two orders of magnitude) than the density expected from the electric field screening. Two other solid samples that were investigated showed similarly low densities of 2.2 and $2.4\times 10^{10}$~m$^{-2}$.    

At temperatures below about 10~K, the resistance is only weakly dependent on $T$, which indicates the scattering of electrons by the surface roughness~\cite{TrojJETP1981,KajitaSS1984,IyeSS1986}. From the estimated value of $n_\textrm{s}/\gamma$, we find the electron mobility corresponding to the resistance $R_\textrm{S}$ to be $\mu=\gamma/(en_\textrm{s}R_\textrm{S})=94$~m$^2$/Vs. Two other solid samples that were investigated showed somewhat lower mobilities of 41 and 48~m$^2$/Vs. This mobility is significantly larger than previously reported. Note that if we used the value of $n_\textrm{s}$ estimated from the electric field screening, as usually assumed, we would obtain a value of $\mu$ due to scattering by the surface roughness to be about two orders of magnitude lower. Thus, a correct estimate of the electron density is an important factor in characterization of the scattering mechanism. Another interesting finding is deviation of the experimental data in the temperature range between 10 and 16~K from the scattering model presented by Eq.~\eqref{eq:R} and the dashed line in Fig.~\ref{fig:7}. In this range, the experimentally found resistance is larger than predicted by an  oversimplified model assuming a temperature-independent resistance $R_\textrm{S}$ due to surface roughness, thus indicating a temperature-dependent mechanism. Other samples that were investigated showed a similar behavior.  

\section{Discussion and conclusion}
\label{sec:disc}

This work presents an experimental study of the electric dc transport of electrons over solid neon and its dependence on neon properties. Despite similar studies have been reported, we faced an unexpected difficulty in interpreting the data in terms of the conventional model describing the experimental method. Nevertheless, we have managed to extract the resistance of the electron system and compare it with the theoretical model. The good agreement of our data with the neon-vapor scattering mechanism that holds at high temperatures allows us to draw some interesting conclusions regarding the system. One of the important results of this work is a surprisingly low density of electrons obtained by charging the surface by thermionic emission from a filament. The density found in our experiment is more than two orders of magnitude lower than usually expected from the electric field screening. A possible reason for this result is charging the bulk of solid neon with high-energy electrons produced by the emission, which penetrate through the surface barrier of about $0.7$~eV. Such electrons, when injected into the bulk of solid neon, become localized and can screen the electrostatic potential due to the positively-biased bottom plate of the setup, thus resulting in a lower density of mobile electrons on the surface. In order to reduce the energy of electrons produced by thermionic emission, a small amount of helium gas can be introduced into the experimental cell to cause the electron energy loss due to collisions with helium atoms~\cite{KajitaSS1982,LeiderPRL1987,Kono2JLTP1991}. In the present work, we avoided adding helium gas to retain purity of the neon system to accurately measure the scattering of electrons by the neon-vapor atoms. We also note that formation of localized electron charge through trapping by the surface roughness of solid neon cannot be excluded~\cite{TrojJETP1979,TrojJETP1981}. Such trapped electrons would not contribute to the dc transport response, but could also lead to the screening of the applied electrostatic potential and reduced density of the mobile electrons.

Another interesting finding is deviation of the electron resistance from the model prediction in the temperature range from 10 to 16 K. This points out a temperature-dependent contribution to the scattering of electrons by surface roughness. According to the kinetic equation method, the mobility for a non-degenerate 2D electron system can be written as~\cite{shikJLTP1974}

\begin{equation}
\mu = \frac{e}{m} \int\limits_0^\infty \frac{x e^{-x}dx}{\nu_\textrm{S}(xk_\textrm{B}T)},
\label{eq:kin}
\end{equation}

\noindent where $m$ is the electron mass and $\nu_\textrm{S}(E)$ is the scattering rate due to surface roughness, which in general can depend on the electron energy $E=p^2/(2m)$. For an energy-independent scattering rate $\nu_\textrm{S}$, the mobility does not depend on temperature. A detailed theoretical analysis of the mobility on solid hydrogen was given by Sokolov {\it et al.} who considered scattering from the static displacements of the solid surface described by a random Gaussian distribution characterized by a correlation length $l$~\cite{SokolovPRB1995}. Depending on the chosen value of $l$, the predicted mobility could show a complicated non-monotonic temperature dependence. For $l\sim 10^{-8}$~m, the mobility showed the temperature dependence $\mu\sim T^{-1}$ at high temperatures above 5~K, which was consistent with the experimental results reported earlier~\cite{TrojJETP1981}. This result is not inconsistent with an increase of resistance for electrons on solid neon observed in our experiment above 10~K. Also, we point out that, even at temperature far below the triple point, drastic structural changes of the noble-gas films can occur~\cite{ClassPRB1995,FleiEur2000,Matk2025}. Thus, a complete theory of scattering by surface roughness requires a microscopic model of surface roughness that would take into account an explicit dependence on temperature. To the best of our knowledge, such a theory does not exist. 

The estimated value of the electron mobility at low temperatures is rather high. In previous experiments, the reported mobility of surface electrons due to surface roughness measured on solid noble-gas substrates did not exceed 10~m$^2$/Vs~\cite{TrojJETP1981,EdelJLTP1983,KajitaSS1984,IyeSS1986,PaalPRL1987}. We note that our analysis is based on the comparison of the experimental data with an oversimplified model of temperature-independent scattering by surface roughness, Eq.~\eqref{eq:R}, and theoretical values for the mobility $\mu$ in the vapor-atom scattering regime. A possible temperature-dependent contribution due to the surface scattering can not be excluded even at high temperatures up to 20~K, which makes the analysis more complicated without knowing a theoretical mechanism. We believe that further improvements of the experimental methods are necessary to confirm the results of this work and investigate the scattering mechanism due to surface roughness in more details. Magneto-resistivity measurements in a magnetic field $B$ applied perpendicular to the surface of solid sample can be used to independently determine the absolute values of the electron density $n_\textrm{s}$ and mobility $\mu$ by virtue of the Drude law for the longitudinal conductivity $\sigma_{xx}=en_\textrm{e}\mu/(1+\mu^2B^2)$. This relation was employed previously to extract information from the magneto-transport data for electrons on liquid helium~\cite{LeaPRB1997} and solid hydrogen~\cite{PaalPRL1987,PaalPRB1988}. However, in light of the results reported here, it is not clear if the electron response on solid neon can be satisfactorily described by conventional models and the values of $n_\textrm{s}$ and $\mu$ can be easily extracted. Other experimental methods, for example ac transport measurements, are desirable towards demonstrating the surface electrons as a sensitive probe of the noble-gas substrate morphology. We hope that the work reported here will be helpful for the analysis of future experiments and contribute to the growing interest in surface electrons on noble-gas solid substrates.       

\begin{acknowledgements}
The work was supported by an internal grant from the Okinawa Institute of Science and Technology (OIST) Graduate University and by Japan's Council for Science, Technology and Innovation (CSTI) under the Cross-ministerial Strategic Innovation Promotional Program (SIP) for "Promoting the application of advanced quantum technology platforms for social issues". We would like to thank Prof. P. Leiderer and Prof. K. Kono for many helpful discussions. 
\end{acknowledgements}

\bibliography{eNeRef}

@article{FleiEur2000,
  title = {{On dewetting dynamics of solid films of hydrogen isotopes and its influence on tritium $\beta$ spectroscopy}},
  author = {Fleischmann, L. and Bonn, J. and Bornschein, B. and Leiderer, P. and Otten, E. W. and Przyrembel, M and Weinheimer, Ch.},
  journal = {Eur. Phys. J. B},
  volume = {16},
  pages = {521--529},
  year = {2000}
}

@article{shikJLTP1974,
	title={On the interaction of surface electrons in liquid helium with oscillations of the vapor-liquid interface},
	author={Shikin, V. B. and Monarkha, Yu. P.},
	journal={J. Low Temp. Phys.},
	volume={16},
	pages={193},
	year={1974},
      publisher={Springer}
}

@article{JinNatPhys2023,
	title = {Electron charge qubit with 0.1 millisecond coherence time},
	author = {Zhou, Xianjing and Li, Xinhao and Chen, Qianfan and Koolstra, Gerwin and Yang, Ge and Dizdar, Brennan and Huang, Yizhong and Wang, Christopher S. and Han, Xu and Zhang, Xufeng and Schuster, David I. and Jin, Dafei},
      volume = {20},
      pages = {116},
	journal = {Nat. Phys.},
	year = {2024},
	doi = {10.1038/s41567-023-02247-5},
}

@article{JinNat2022,
	title = {Single electrons on solid neon as a solid-state qubit platform},
	author = {Zhou, Xianjing and Koolstra, Gerwin and Zhang, Xufeng and Yang, Ge and Han, Xu and Dizdar, Brennan and Li, Xinhao and Divan, Ralu and Guo, Wei and Murch, Kater W. and Schuster, David I. and Jin, Dafei},
	volume = {605},
	number = {7908},
	urldate = {2023-06-21},
	journal = {Nature},
	year = {2022},
	pages = {46},
	doi = {10.1038/s41586-022-04539-x},
}

@article{JinNatElec2026,
	title={{Solid neon as a noise-resilient host for electron qubits above 100 mK}},
	author={Li, X. and Wang. C. S. and Dizdar, B. and Huang, Y. and Wen, Y. and Guo, W. and Zhang, X. and Han, X. and Zhou, X. and Jin. D},
	journal={Nature Electronics},
	volume={9},
	pages={621},
	year={2026},
	publisher={Nature}
}

@article{KanaiPRL2024,
      title={Single-electron qubits based on ring-shaped surface states on solid neon},
      author={Toshiaki Kanai and Dafei Jin and Wei Guo},
	journal={Phys. Rev. Lett.},
	volume={132},
	pages={025603},
      year={2024}
}

@article{Chen2022,
	author = {Chen, Qianfan and Martin, Ivar and Jiang, Liang and Jin, Dafei},
	journal={Quantum Sci. and Technol.},
	volume = {7},
	pages = {045016},
	title = {{Electron spin coherence on a solid neon surface}},
	year = {2022},
	publisher={IOP}
}

@article{Zou2026,
	author = {Zou, S. and Li,X. and Chen, Q. and Jin, D},
	journal={Quantum Sci. and Technol.},
	volume = {11},
	pages = {025003},
	title = {{Electron charge coherence on a solid neon surface}},
	year = {2026},
	publisher={IOP}
}

@article{LeiderJLTP2025rew,
	title={Surface electrons on solid quantum substrates: a brief review},
	author={Leiderer, Paul},
	journal={J. Low Temp. Phys.},
	volume={219},
	pages={262--281},
	year={2025},
	publisher={Springer}
}

@article{SokolovPRB1995,
 	 title = {Theory of electron transport in a two-dimensional layer on the solid hydrogen surface},
  		author = {Sokolov, Sviatoslav S. and Rino, Jos\'e Pedro and Studart, Nelson},
 	 journal = {Phys. Rev. B},
 	 volume = {51},
  	pages = {11068--11073},
  	year = {1995},
  	publisher = {American Physical Society},
  	doi = {10.1103/PhysRevB.51.11068}
}

@article{SaitohJPSJ1977,
  	title = {{Warm electrons on the liquid $^4$He surface}},
  	author = {Saitoh, M},
  	journal = {J. Phys. Soc. Jpn.},
  	volume = {42},
  	pages = {201--209},
  	year = {1977}
}

@article{Kono1JLTP1991,
  	title = {{Surface-state electrons on a hydrogen film. 1. Annealing of the film}},
  	author = {Kono, K. and Albrecht, U. and Leiderer, P.},
  	journal = {J. Low Temp. Phys.},
  	volume = {82},
  	pages = {279},
  	year = {1991}
}

@article{Kono2JLTP1991,
  	title = {{Surface-state electrons on a hydrogen film. 2. Influence of absobed helium films}},
  	author = {Kono, K. and Albrecht, U. and Leiderer, P.},
  	journal = {J. Low Temp. Phys.},
  	volume = {83},
  	pages = {423},
  	year = {1991}
}

@article{TrojJETP1979,
	title = {Electron localization over the surface of crystallie hydrogen and neon},
	author = {Troyanovskii, A. M. and  Volodin, A. P. and Khaikin, M. S. },
	journal={JETP Lett.},
	volume = {29},
	pages = {382},
	year = {1979}
}

@article{TrojJETP1981,
	title = {Electron mobility in two-dimensional layer over the surface of solid hydrogen},
	author = {Troyanovskii, A. M. and Khaikin, M. S. },
	journal={Sov. Phys. JETP},
	volume = {54},
	pages = {214},
	year = {1981}
}

@article{KajitaSS1984,
	title = {A new two-dimensional electron system on the surface of solid neon},
	author = {Kajita, K.},
	journal = {Surf. Sci.},
	pages = {86},
	volume = {142},
	year = {1984}
}

@article{KajitaSS1986,
	title = {{Electrical conductivity of highly correlated 2D electrons formed on solid neon}},
	author = {Kajita, K. and Nishio, Y. and Sasaki, W.},
	journal = {Surf. Sci.},
	pages = {88},
	volume = {170},
	year = {1986}
}

@article{KajitaSS1982,
	title = {Two-dimensional electrons on the helium film-solid neon system},
	author = {Kajita, K. and Sasaki, W.},
	journal = {Surf. Sci.},
	pages = {419},
	volume = {113},
	year = {1982}
}

@article{IyeSS1986,
	title = {Mobile electrons on a solid hydrogen surface},
	author = {Paalanen, M. A. and Iye, Y},
	journal = {Surf. Sci.},
	pages = {80},
	volume = {170},
	year = {1986}
}

@article{IyeJLTP1980,
  	title = {Mobility of electrosn in the surface state of liquid helium},
  	author = {Iye, Y.},
  	journal = {J. Low Temp. Phys.},
  	volume = {40},
  	pages = {441},
  	year = {1980}
}

@article{EdelJLTP1983,
  	title = {Invetsigation of electrons localized above solid hydrogen by means of the cyclotron resonace method},
  	author = {Edel'man, V. S. and Faley, M. I.},
  	journal = {J. Low Temp. Phys.},
  	volume = {52},
  	pages = {301},
  	year = {1983}
}

@article{PaalPRL1987,
	title={Localization in a nondegenerate two-dimensional electron gas},
	author={Adams, P. W. and Paalanen, M. A.},
	journal={Phys. Rev. Lett.},
	volume={58},
      number={20},
	pages={2106},
	year={1987},
	publisher={APS}
}

@article{PaalPRB1988,
	title={High-field magnetoconductivity of electrons on hydrogen},
	author={Adams, P. W. and Paalanen, M. A.},
	journal={Phys. Rev. B},
	volume={37},
      number={7},
	pages={3805},
	year={1988},
	publisher={APS}
}

@article{LeiderPRL1987,
	title={Invetsigation of thin helium films with surface-bound electrons},
	author={Cieslikowski, D. and Dahm, A. J. and Leiderer, P.},
	journal={Phys. Rev. Lett.},
	volume={58},
      number={17},
	pages={1751},
	year={1987},
	publisher={APS}
}

@article{SommerPRL1971tanner,
	title={{Mobility of electrons on the surface of liquid $^4$He}},
	author={Sommer, W. T. and Tanner, David J},
	journal={Phys. Rev. Lett.},
	volume={27},
	number={20},
	pages={1345},
	year={1971},
	publisher={APS}
}

@article{WilenJLTP1988,
	title={Impedance method for surface state electrons},
	author={Wilen, L. and Giannetta, R.},
	journal={J. Low Temp. Phys.},
	volume={72},
	pages={353},
	year={1988},
      publisher={Springer}
}

@article{DahmJLTP1987,
	title={{Analisis of the Sommer technique for measurement of the mobility for chrages in two dimensions}},
	author={Mehrotra, R. and Dahm, A. J.},
	journal={J. Low Temp. Phys.},
	volume={67},
	pages={115},
	year={1987},
      publisher={Springer}
}

@article{KovdJLTP1993,
	title={Quantum transport of surface electrons over liquid helium in magnetic fields},
	author={Kovdrya, Yu. Z. and Nikolaenko, V. A. and Kirichek, O. I. and Sokolov, S. S. and Grigor'ev, S. S. },
	journal={J. Low Temp. Phys.},
	volume={91},
	pages={371},
	year={1993},
      publisher={Springer}
}

@article{SyvokLTP1996,
	title={Nonlinear phenomena in the mobility of two-dimensional electron layer at the liquid helium surface},
	author={Sivokon, V. E. and Dotsenko, V. V. and Sokolov, S. S. and Kovdrya, Yu. Z. and Grigor'ev, V. N.},
	journal={Low Temp. Phys.},
	volume={22},
	pages={549},
	year={1996}
}

@article{HarrisPRA1970,
	title={{Thermal expansion of liquid helium II}},
	author={Harris-Lower, R. F. and Smee, K. A.},
	journal={Phys. Rev. A},
	volume={2},
      number={1},
	pages={158},
	year={1970},
	publisher={APS}
}

@article{EbnPRA1979,
  title = {{Density-functional theory of simple classical fluids. II. Localized excess electron states}},
  author = {Ebner, C. and Punyanitya, C.},
  journal = {Phys. Rev. A},
  volume = {19},
  issue = {2},
  pages = {856--865},
  year = {1979},
  publisher = {American Physical Society}
}

@article{AcoPRB2000,
  title = {Thermal and elastic properties of solid neon},
  author = {Acocella, D. and Horton, G. K. and Cowley, E. R.},
  journal = {Phys. Rev. B},
  volume = {61},
  issue = {13},
  pages = {8753},
  year = {2000},
  publisher = {American Physical Society}
}

@article{ClassPRB1995,
	title={Resonant scattering of surface acoustic waves by hydrogen crystallites},
	author={{Classe, J. and Eschenr\:oder, K. and Weiss, G.}},
	journal={Phys. Rev. B},
	volume={52},
	pages={11475},
	year={1995},
	publisher={APS}
}

@article{MurchPRL2025,
	title={Surface-Morphology-Assisted Trapping of Strongly Coupled Electron-on-Neon Charge States},
	author={Zheng, K. and Song. X. and Murch, K. W.},
	journal={Phys. Rev. Lett.},
	volume={135},
	pages={080601},
	year={2025},
	publisher={APS}
}

@article{MurchPRAppl2026,
	title={Real-time monitoring of growth of neon film for electrosn-on-neon qubits},
	author={Duthaluru, S. and Zheng, K. and Henriksen, E. A. and Murch, K. W.},
	journal={Phys. Rev. Applied},
	volume={25},
	pages={044065},
	year={2026},
	publisher={APS}
}

@misc{Matk2025,
      title={{Characterizing Neon Thin Film Growth with an NbTiN Superconducting Resonator Array}},
      author={Matkovic, K. and Russell, P. and Palmer, A. and Helgemo, E. and Delventhal, L. and Zuo, K. and Surse, K. and Rahman, R. and Cassidy, M. C.},
      year={2025},
      eprint={2510.21029},
      archivePrefix={arXiv},
      primaryClass={cond-mat.mes-hall}
}

@misc{NIST,
	title={{https://webbook.nist.gov/cgi/inchi/InChI\%3D1S/Ne}},
}

@article{leaPRB1997,
	title={Magnetoconductivity of two-dimensional electrons on liquid helium: experiments in the fluid phase},
	author={Lea, M. J. and Fozooni, P. and Kristensen, A. and Richradson, P. J. and Djerfi, K. and Dykman, M. I. and Fang-Ye, C. and Blackburn, A},
	journal={Phys. Rev. B},
	volume={55},
	pages={16280},
	year={1997},
	publisher={APS}
}

@misc{Dizd2026,
      title={{Probing wetting and solidification dynamics of neon thin films with a high-$T_c$ superconducting circuit}},
      author={Dizdar, Brennan and Shi, Yueheng and  Schuster, David I. and Lang, Christopher S.},
      year={2026},
      eprint={2610.24167},
      archivePrefix={arXiv},
      primaryClass={cond-mat.mes-hall}
}

\end{document}